\RequirePackage{fixltx2e}
\documentclass[floatfix,aip,jcp,preprint,showkeys]{revtex4-1} 
\usepackage{graphicx}
\usepackage{mathrsfs}
\usepackage{gensymb}
\usepackage{amssymb}
\usepackage{amsmath}
\usepackage{textcomp}
\usepackage{dblfloatfix}
\usepackage{mhchem}%
\usepackage{hyperref}
\usepackage{array}
\usepackage{booktabs}
\usepackage{multirow}
\usepackage{dcolumn}
\usepackage[caption=false,font=footnotesize]{subfig}
\AtBeginDocument{
  \heavyrulewidth=.08em
  \lightrulewidth=.05em
  \cmidrulewidth=.03em
  \belowrulesep=.65ex
  \belowbottomsep=0pt
  \aboverulesep=.4ex
  \abovetopsep=0pt
  \cmidrulesep=\doublerulesep
  \cmidrulekern=.5em
  \defaultaddspace=.5em
}

\begin{document}
\title{A stochastic model of solid state thin film deposition: application to chalcopyrite growth}
\date{\today}
\author{Robert J. \surname{Lovelett}}
\author{Xueqi \surname{Pang}}
\author{Tyler M. \surname{Roberts}}\affiliation{Department of Chemical and Biomolecular Engineering, University of Delaware, Newark, DE 19716}\affiliation{Institute of Energy Conversion, University of Delaware, Newark, DE 19716}
\author{William N. \surname{Shafarman}}
\author{Robert W. \surname{Birkmire}}\affiliation{Department of Materials Science and Engineering, University of Delaware, Newark, DE 19716}\affiliation{Institute of Energy Conversion, University of Delaware, Newark, DE 19716}
\author{Babatunde A. \surname{Ogunnaike}}\email[E-mail: ]{ogunnaike@udel.edu}\affiliation{Department of Chemical and Biomolecular Engineering, University of Delaware, Newark, DE 19716}
\begin{abstract}
	Developing high fidelity quantitative models of solid state reaction systems can be challenging, especially in deposition systems where, in addition to the multiple competing processes occurring simultaneously, the solid interacts with its atmosphere.
	In this work, we develop a model for the growth of a thin solid film where species from the atmosphere adsorb, diffuse, and react with the film.
	The model is mesoscale and describes an entire film with thickness on the order of microns.
	Because it is stochastic, the model allows us to examine inhomogeneities and agglomerations that would be impossible to characterize with deterministic methods.
        We demonstrate the modeling approach with the example of chalcopyrite Cu(InGa)(SeS)\textsubscript{2} thin film growth via precursor reaction, which is a common industrial method for fabricating thin film photovoltaic modules.
	The model is used to understand how and why through-film variation in the composition of Cu(InGa)(SeS)\textsubscript{2} thin films arises and persists.
	We believe that the model will be valuable as an effective quantitative description of many other materials systems used in semiconductors, energy storage, and other fast-growing industries.
\end{abstract}
\maketitle
\section{Introduction}
Quantitative understanding of solid state reactions involved in film deposition and growth is important for improved processing in a number of industries including microelectronics, photovoltaics, energy storage, and pharmaceuticals.
While there are many useful classical modeling approaches (from simple mass action kinetics \cite{Verma1996, Orbey1997} to detailed Avrami-type modeling \cite{Avrami1939, Avrami1941, Kim2005, Purwins2006}) the increasingly complex material systems used in modern manufacturing require more sophisticated methods.
Most modern techniques, however, employ microscopic level or \textit{ab initio} approaches (\textit{e.g.}, Refs.\@ \onlinecite{Wimmer2000, Ryzhikov2015}), which promote fundamental knowledge but, because of current computational hardware limitations, are incapable of providing larger-scale property or composition predictions.
In this paper, we present a mesoscopic thin film growth model capable of predicting film-scale composition. 

In order to examine and characterize the lateral heterogeneity that can arise during film growth, the model we present is stochastic, rather than deterministic.
Our approach is related to the stochastic simulation algorithm first developed by Gillespie \cite{Gillespie1977}, which assumes a uniformly mixed system with no mass transfer limitations.
In solid state systems, however, mass transfer effects are always important and often rate-limiting.
Gillespie's method has recently been extended to include diffusion, mostly used for micro-scale modeling of biomolecular systems.
Erban and Chapman \cite{Erban2009} discuss two general approaches: on-lattice methods and off-lattice methods.
On-lattice approaches restrict the position of molecules to discrete locations or compartments where each compartment contains multiple occupant species, while off-lattice methods allow movement on a continuous domain, usually through Brownian motion.
However, these approaches are designed for fluid systems where species density may vary.
In this work, we apply similar concepts and expand the capability of stochastic models for reactions in the solid state.
Our approach, discussed in Section \ref{sec:approach}, is similar to on-lattice methods, but with the additional restriction that lattice occupancy is always exactly one.
Instead of interacting with co-occupants, species interact with adjacent lattice points.
We show how this approach allows stochastic simulation of mesoscale systems of solid, crystalline species, where unit-cell level (1 \AA) simulation would be impractical for complete thin film (1--10~\micro m) systems.
Although we assume the lattice is square, with a coordination number of four, our approach is easily generalized to allow for simulation of advanced, non-isotropic materials such as graphene, carbon nanotubes, and other materials with complex microstructure.
We use the chalcopyrite Cu(InGa)(SeS)\textsubscript{2} system as an example to demonstrate how one can simulate film growth rate, composition profile, and agglomeration using the proposed stochastic approach.

Polycrystalline chalcopyrite CuInSe\textsubscript{2}-based materials are commonly used as the absorber layer in thin film solar cells.
Devices using these materials have demonstrated efficiencies exceeding 20\% \cite{Jackson2011, chirila2013, SolarFrontier2014}.%
In order to increase voltage and improve efficiency, gallium and sulfur are alloyed with CuInSe\textsubscript{2} to form a continuous solid solution: Cu(InGa)(SeS)\textsubscript{2}.
The most common industrial process for producing these absorbers involves two steps in which a metal precursor (Cu-In-Ga) is deposited first and then reacted with gas-phase H\textsubscript{2}Se and/or H\textsubscript{2}S \cite{Shafarman2011}.
The reaction step can result in heterogeneous films with steep through-film composition gradients \cite{Jensen1993} and spatially-confined agglomerations \cite{Hanket2007, Kim2012}.
The spatial heterogeneity resulting from the selenium and sulfur reactions makes this process an ideal system to demonstrate our solid state reaction model.

In the remaining sections, we present a novel stochastic model for solid state reaction kinetics, with emphasis on the ability to predict the composition profile and other spatial heterogeneities.
In Section \ref{sec:approach} we develop the model, describe an efficient solution algorithm, and explain the relationship between the model parameters and physical properties.
Then, in Section \ref{sec:CIGS} we describe a reaction mechanism for Cu(InGa)(SeS)\textsubscript{2} production, show how to apply the model using this mechanism, use the model to predict composition profiles and agglomeration statistics, and compare model predictions and experimental results.
Finally, we offer conclusions and suggestions for future applications.

\section{Model Development and Theory}
\label{sec:approach}
The system in question is a thin film in which solid state reactions occur, species interdiffuse, and the film interacts with its environment by adsorption and desorption of volatile species.
We represent the film with a two-dimensional square lattice, where each point contains a species or a vacancy.
The model is mesoscopic; so that each lattice point does not represent an individual atom, molecule, or unit cell, otherwise the lattice would be too large to be computationally tractable.
The lattice is therefore a coarse-grained approximation of the actual film; each lattice point is a finite volume element small enough such that it is accurately approximated as phase-pure.

Our approach is to recast Gillespie's stochastic simulation algorithm \cite{Gillespie1977} for spatially heterogeneous solid state systems with approximately constant mass density and number density.
In Gillespie's method, a random number is selected at each time step to determine which reaction occurs.
Here, we generalize reaction events to ``lattice" events, which take place at interfaces between lattice points and are classified as reaction, diffusion, adsorption, or desorption events.
The probability and the rate of occurrence of each lattice event are governed by an intrinsic parameter called the propensity constant.
The propensity of a given event is the product of its propensity constant and the number of interfaces at which the event can occur.

The modeling approach is as follows:
\begin{enumerate}
	\item A square lattice is initialized with the starting species. If adsorption/desorption events are included, the lattice should contain vacancy points above the species. If the lattice is represented as an $N \times M$ array, row 0 and row $N$ are considered boundaries with no interactions above row 0 points or below  row $N$ points. Column 0 is considered adjacent with column $M$ (\textit{cf.}, periodic boundary conditions in a boundary value problem involving a partial differential equation).
	\item Propensity of each lattice event is calculated as: $a_i=p_iN_{i}$, where $p_i$ is the propensity constant and $N_{i}$ the number of interfaces associated with lattice event $i$.
        \item Probability of each lattice event is proportional to the propensity of an event; a random number is generated to determine which lattice event occurs.
        \item The time, $\tau$, until the next time step is selected from an exponentially distributed random variable.
        \item The reaction chosen in Step~3 occurs at one possible interface. For example, if a reaction takes place between species A and B, then one of the A---B interfaces is selected at random and updated accordingly. For reaction events, the final orientation (that is, the relative position of the product species) is random; it is fixed for adsorption, and diffusion events.
	\item The lattice is updated and steps 2--5 are repeated until an exit condition is met.
\end{enumerate}

\subsection{Solution Algorithm}
\label{sec:algo}
The conceptually simplest algorithm for implementing this modeling approach is to store the lattice in a 2D array that is updated at each time step.
Although straightforward, this method is inefficient and will be computationally tractable only for relatively small lattice sizes.
There are two possible bottlenecks that can lead to very slow solutions: (1) counting the number of interfaces of each kind, and (2) choosing one of these interfaces at random for a reaction event.
To address these issues, we do not store and update the lattice itself at each time step; instead we track each interface and its position in an array.
First, we define the simple 2D representation of the lattice; then we demonstrate how to convert this to the 1D representation that is used in the algorithm.

Consider a film discretized to a lattice and represented by the 2D array: $\mathbf{L}\in\mathbb{N}_0^{(N\times M)}$ with elements $l_{(i,j)}$.
The indices of each element in $\mathbf{L}$ represent the position of each lattice point (\textit{i.e.}, volume element) in physical space, and the value of that element represents its occupant species.
If there are $S$ unique species, including vacant elements, and the value of each array element corresponds to its occupant species, then the domain of $l_{(i,j)}$ is given as: $\{l_{(i,j)} |l_{(i,j)} \in \mathbb{N}_0, l_{(i,j)} < S\}$.
Now, consider the set of adjacent points in $\mathbf{L}$.
For this model, we define the set of adjacent elements to be: 
\begin{equation}
  \label{eqn:adj_def}
(l_{\textrm{adj}}^1,l_{\textrm{adj}}^2) | (l_{\textrm{adj}}^1,l_{\textrm{adj}}^2) \in \begin{cases} (l_{(i,j)}, l_{(i,j-1)}) \forall j>0,\\ (l_{(i,j)},l_{(i-1,j)})\forall i>0, \\ (l_{(i,0)}, l_{(i,M-1)})\end{cases}
\end{equation}
where the first two cases are trivially adjacent points and the third is analogous to applying periodic boundary conditions to a PDE to reduce edge effects.
The periodic boundary condition is applied only to the columns of $\mathbf{L}$, as the rows represent the full thickness of the thin film.
Next, we convert the 2D representation, $\mathbf{L}$, to a 1D representation $\mathbf{X}$, with elements $x_k$ by: (1) defining a mapping from \textit{species pairs} to \textit{interface kinds}, or, $x = f(l_{\textrm{adj}}^1,l_{\textrm{adj}}^2)$, and (2) define a mapping from indices in $\mathbf{L}$ to index in $\mathbf{X}$, or $k = g((i_1,j_1),(i_2,j_2))$, where $((i_1,j_1), (i_2,j_2))$ are the indices of $(l_{\textrm{adj}}^1,l_{\textrm{adj}}^2)$.

Each adjacent pair of lattice elements constitutes an interface, which can be represented by a single value $\{x|x\in\mathbb{N}_0, x<S^2\}$, where there are $S^2$ ``kinds" of interfaces (observe that interfaces of different orientation are considered distinct).
We can then map \textit{pairs of species} to \textit{interface kinds}:
\begin{equation}
  \label{eqn:spec_to_int}
  x = l_{\textrm{adj}}^1S+l_{\textrm{adj}}^2
\end{equation}
where $x$ is the interface kind.

Next, we define a mapping for the indices for interface location in $\mathbf{L}$ to the index for interface location in $\mathbf{X}$.
Here, we assume that the number of columns, $M$, is an even number (a similar procedure applies if $M$ were odd-valued).
With $((i_1,j_1),(i_2,j_2))$ as indices of adjacent elements in $\mathbf{L}$ and $k$ as the index of $\mathbf{X}$:
\begin{equation}
	\label{eqn:pos_index}
	k = \begin{cases} j_1 & \textrm{if }  i_1=i_2=0 \\
	                  M + 2M(i_1-1) + 2j_1 & \textrm{if }  i_1=i_2\ne0\\
	                  M + 2M(i_1-1) + 2j_1 + 1   & \textrm{if }  j_1=j_2
	    \end{cases}
\end{equation}

Using the mappings defined in Equations \ref{eqn:spec_to_int} and \ref{eqn:pos_index} to translate $\mathbf{L}\rightarrow\mathbf{X}$, the simulation algorithm may now be written as follows:
\begin{enumerate}
	\item Define a set of allowable species, $\{0, 1, 2,\ldots (S-1)\}$. 
        \item Define a set of directional lattice events, $\mathbf{D}$. Directional events will preserve the orientation of the interface and are used to represent diffusion, adsorption, and desorption events. $\mathbf{D}$ is a ($N_D \times 3$) matrix, where $N_D$ is the number of directional lattice events. The columns of $\mathbf{D}$ correspond to $\{p, x_0, x_f\}$ for the propensity constant, initial interface kind, and final interface kind, respectively.
        \item Define a set of non-directional lattice events, $\mathbf{N}$. Non-directional events will not preserve the orientation of the interface and are used to represent reaction events. $\mathbf{N}$ is a  ($3 \times N_N$) matrix, where $N_N$ is the number of non-directional lattice events. The columns of $\mathbf{N}$ correspond to $\{p, x_0, x_f\}$ for the propensity constant, initial interface kind, and final interface kind, respectively. Directionality refers to the orientation of the \textit{products}, not the reactants. For example, if there are species A, B, C, and D, then the reaction \ce{A + B -> C + D} is distinct from \ce{B + A -> C + D}.
	\item Define an initial condition $\mathbf{L_0}$.
	\item Use Equations \ref{eqn:spec_to_int} and \ref{eqn:pos_index} to map $\mathbf{L_0}\rightarrow\mathbf{X}$
	\item Count the number of interfaces of each kind in array $\mathbf{X}$, saving the results in array $\mathbf{Y}$, referred to as the ``interface count array":
		\begin{equation}
			\label{eqn:interface_count}
			y_m = \sum_{m \in \mathbf{X}} 1, \textrm{ for } m\in\mathbb{N}_0, m<S^2
		\end{equation}
        \item Calculate the total propensity array, $\mathbf{A}\in\mathbb{R}^{((\textrm{rows}(\mathbf{D})+\textrm{rows}(\mathbf{N}))\times 1)}$ from each event in $\mathbf{D}$ and $\mathbf{N}$, which has the elements:
		\begin{equation}
			\label{eqn:event_total_prop}
                        \begin{split}
                          a_i &= \{p\in\mathbf{D}_i\} y_{x_0\in\mathbf{D}_i} \textrm{ and} \\
                          a_{i+\textrm{rows}(\mathbf{D})} &= \{p\in\mathbf{N}_i\} y_{x_0\in\mathbf{N}_i}
                        \end{split}
		\end{equation}
	\item Determine the time step, $\tau$, drawn from the probability mass function:
		\begin{equation}
			\label{eqn:tau_pdf}
                        Pr(\tau|\mathbf{A})=\sum a_i \exp{(\tau \sum a_i)}
		\end{equation}
	\item Determine which lattice event occurs using probabilities:
		\begin{equation}
			\label{eqn:event_probability}
			P_i = \frac{a_i}{\sum a_j}
		\end{equation}
	\item From a uniformly random distribution, choose the reactive interface, \textit{i.e.}, the allowable interface at which the reaction takes place. 
        \item Update $\mathbf{X}$ and $\mathbf{Y}$. The reactive interface (an element in $\mathbf{X}$) and the interfaces adjacent to the reactive interface will be updated. The interface count array $\mathbf{Y}$ is updated by subtracting the previous the values of $\mathbf{X}$  and adding the new values of $\mathbf{X}$ to their corresponding elements in $\mathbf{Y}$. (Refer to the code at the URL below for a complete definition of adjacent elements in $\mathbf{X}$).
        \item Check if exit condition is reached. If yes: continue; else: return to Step 7. The exit condition should be determined on a case-by-case basis. In this work, a specified simulation time is used and alternatives include a specified number of time steps or composition (though specifying a composition as an exit condition will not ensure that the composition will ever be reached).
        \item Using the inverses of Equations  \ref{eqn:spec_to_int} and \ref{eqn:pos_index}, calculate $\mathbf{L}$. End.
\end{enumerate}

It should now be clear that the solution of the model is equivalent to sampling from a Markov chain where the elements of $\mathbf{X}$ define the system state; further, each step of the Markov chain will be assigned a step time in continuous time by assuming that each step is a Poisson process with intensity equal to the sum of all event propensities.
The algorithm has been written in Python using Numpy \cite{vanderwalt2011} and the code is available at \url{www.bitbucket.org/rlovelett/stochastic_solid_state}. Moderately sized simulations (about 1000 elements in $\mathbf{L}$, 100 elements in $\mathbf{D}$ and $\mathbf{N}$, and $10^7$ time steps) run in several minutes on a typical personal computer.

\subsection{Relation to Physical Properties}
\label{sec:properties}
One of the advantages of a spatially distributed model is its ability to decouple the rates of different processes---reactions, adsorption, desorption, and diffusion.
The physical properties that govern these rates can be related to the propensity constants used to advance the model.
We now show how to derive physical property values from the value of the propensity constants.
\subsubsection{Reaction Propensity}
In its original implementation, the Gillespie algorithm's propensity constant, $c_i$, is related to the macroscopic reaction rate constant, $k_i$ according to $k_i=V^{b-1}c_i$, where $V$ is the reactor volume and $b$ is the reaction order.
Our system is analogous, except that our reactions occur at surfaces.
We assume that every formula unit on a lattice site is available for reaction at the surface, \textit{i.e.}, that diffusion within a lattice site is rapid.
The characteristic surface area for reaction is $\ell^2$, where $\ell$ is the length of a lattice site.
Therefore, we can write the rate constant as:
\begin{equation}
  \label{eqn_reaction}
  k_i=\ell^2\rho_{n,s}p_i
\end{equation}
where $\rho_{n,s}$ is the area specific number density (units of inverse area) of formula units.
With enough data at varying temperature, activation energies of the reactions can be estimated using the Arrhenius equation.
\subsubsection{Adsorption Propensity}
The rate of adsorption can be described by the product of the rate of collisions between a surface and a gas, and a sticking coefficient, $S_i$, or the probability that the species will stay on the surface.
In some cases, like the chalcopyrite growth system we examine in Section \ref{sec:CIGS}, the probability of dissociation of a gas phase species should also be included.
For simplicity, however, probability of hydride gas dissociation will be lumped with the sticking coefficient.
First, from kinetic theory, we can determine the rate of collisions with the surface per area, $F_{i,a}=P_i/\sqrt{2\pi M_ik_bT}$, where $F_{i,a}$ is the adsorptive flux of species~$i$ (m$^{-2}$s$^{-1}$); $P_i$, the partial pressure; $M_i$, the molecular weight; $k_b$, Boltzmann's constant; and $T$, the absolute temperature.
Therefore, we can determine the rate of adsorption on a single lattice site using the area of that lattice site: $r_{ads}=F_{i,a}S\ell^2$.
However, this rate is in \textit{molecules} per time, not \textit{lattice sites} per time; thus, it should be scaled by the number of molecules in a lattice site, or $\rho_n\ell^3$, where $\rho_n$ is the number density of the species.
Since $1/p_i$ is the average time until an adsorption event occurs on a single site, $p_i$ is the stochastic equivalent of the rate of adsorption on a single lattice site.
If the propensity constant is known, then we can derive an expression for $S_i$ as:
\begin{equation}
  \label{eqn_adsorption}
  S_i=\frac{p_i\sqrt{2\pi M_ik_bT}}{\rho_n \ell^5P_i}
\end{equation}
\subsubsection{Desorption Propensity}
Desorption is physically equivalent to evaporation.
Unlike other lattice events, however, the rate of evaporation is not a function of thermodynamic properties only; it depends also on system-specific parameters such as gas flow rate and reactor geometry.
Therefore, the rate of evaporation is best captured by a mass transfer coefficient, $k_{m,i}=F_{i,e}/{\Delta f_i}$, where $F_{i,e}$ is the evaporative flux, and $\Delta f_i$ is the difference in the species $i$ fugacity between the adsorbed phase and gas phase.
Assuming that the gas phase is ideal, the fugacity difference reduces to $(P_{vap,i}-P_i)$.
Similar to the adsorption case, we can determine the rate of evaporation from a single lattice site, $r_{evap}=F_{i,e}\ell^2$, scale it by the number of molecules in a lattice site, and set the scaled evaporation rate equal to the propensity constant.
Solving for $k_{m,i}$ yields:
\begin{equation}
  \label{eqn_desorption}
  k_{m,i}=\frac{p_i}{\rho_n\ell^5(P_{vap,i}-P_i)}
\end{equation}

\subsubsection{Diffusion Propensity}
Multicomponent systems with significant diffusion limitations can be challenging to model appropriately.
However, because our modeling approach involves binary interactions between species, we can relate the propensity constants for diffusion events to binary diffusivities.
We will use a well-known result from statistical physics to obtain the diffusivities, where diffusivity can be written as:
\begin{equation}
  \label{eqn_diff_auto}
  D_{s_1,s_2}=\int_t^\infty R(t') dt'
\end{equation}
Here $D_{s_1,s_2}$ is the diffusivity and $R(t')$ is the velocity autocorrelation function of species $s_1$ surrounded by species $s_2$, defined as $R(t')=\langle\mathbf{v}(t)\cdot\mathbf{v}(t+t')\rangle$, denoting the inner product of $\mathbf{v}(t)$ and $\mathbf{v}(t+t')$.
In our system, consider a single lattice point with value $s_1$ surrounded by an infinite lattice species $s_2$.
In this case, because there are four $s_1,s_2$ interfaces, the propensity for diffusion is $a_{s_1,s_2}=4p_{s_1,s_2}$, which means that the average time until the occurrence of a diffusion event is:
\begin{equation}
  \label{eqn_diff_rate}
   \overline{\Delta t_k}=\frac{1}{4p_{s_1,s_2}}
\end{equation}
Therefore, the velocity magnitude of species $i$ is $4p_{s_1,s_2}\ell$.
Choosing the current time to be immediately before a diffusion event, and recognizing that our model consists of discrete time steps, we can rewrite Equation \ref{eqn_diff_auto} as:
\begin{equation}
  \label{eqn_diff_auto_as_sum}
  D_{s_1,s_2}= \langle\mathbf{v_0}\cdot\mathbf{v_0}\rangle \Delta t_0 + \sum_{k=1}^\infty \langle\mathbf{v_0}\cdot\mathbf{v_k}\rangle \Delta t_k
\end{equation}
where $\mathbf{v}_k$ is the velocity of species $s_1$ during time step $k$, $\Delta t_k$ is the duration of time step $k$, and initial element of the series ($k=0$) is moved outside the summation.
Considering that the direction of the velocity vector $\mathbf{v_k}$ will be chosen randomly from two orthogonal unit vectors and their inverses at each time step, we conclude that the sum will converge to zero, and that only the average value of the first term should be retained.
Therefore, by using the average velocity magnitude and average time until the occurrence of an event given in Equation \ref{eqn_diff_rate}, the binary diffusivity is obtained as:
\begin{equation}
  \label{eqn_diffusivity}
  D_{s_1,s_2} = 4p_{s_1,s_2}\ell^2
\end{equation}

Finally, we should note that our approach is unconventional for describing diffusion in solids.
Typically, the crystallinity of the material should be taken into account explicitly and non-isotropic effects should be considered.
Our analysis, however, does not include non-isotropic effects or explicitly consider the presence of discrete crystalline grains.
Therefore, our approach will strictly only be valid if (1) the crystal does not show preferential orientation \textit{and} (2) grain interior diffusion (such as interstitial or vacancy-mediated diffusion) occurs at a rate similar to or greater than grain boundary diffusion.
The second condition could be very restrictive, as grain boundary diffusion usually dominates in polycrystalline films.
However, if this assumption were to be violated, the result would be that the diffusion propensities in the model would correspond to \textit{effective} diffusion coefficients (via Equation \ref{eqn_diffusivity}).
In this case, the diffusion coefficient would not correspond to the energy of a fundamental reaction step or obey the Arrhenius relation that is typical of solid state diffusivities.

\section{Application to C\lowercase{u}(I\lowercase{n}G\lowercase{a})(S\lowercase{e}S)\textsubscript{2} film growth}
\label{sec:CIGS}
\subsection{Reaction Mechanism}
To apply the model to the reaction of Cu-In-Ga precursors with H\textsubscript{2}Se and H\textsubscript{2}S, we require a reaction mechanism.
Several groups \cite{Jensen1993, Purwins2006, Hergert2006, Koo2011} have suggested plausible reaction pathways.
The specific phases or species involved vary somewhat among groups, suggesting that the reaction pathway may be process dependent.
However, some elements are common to each mechanism.
First, there are at least two stages to the reaction: (1) metal chalcogenide formation (\textit{e.g.}, InSe, InS, Ga\textsubscript{2}Se\textsubscript{3}), and (2) chalcopyrite formation (CuInSe\textsubscript{2}, CuInS\textsubscript{2}, etc.).
Second, although reaction rates are often not determined quantitatively, it is suggested that the reaction of Se and In is faster than the reaction of Se and Ga \cite{Hanket2007}.
This asymmetry in reaction rates leads to mostly CuInSe\textsubscript{2} near the front of the film and Ga-containing species accumulated at the back.
Next, we note that the reaction mechanism we propose implies that the Cu(InGa)(SeS)\textsubscript{2} films are stoichiometric and that the discrete nature of the model enforces the stoichiometry.
In practice, however, the precursors (and consequently the final films) are deposited as copper-deficient with the ratio Cu/(In+Ga) $\approx$ 0.9 \cite{Shafarman2011, Purwins2007}.
We do not treat copper deficiency explicitly, though the diffusion coefficients (and, in this model, diffusion propensities) would change if stoichiometric or copper rich-film films were modeled because indium and gallium may diffuse through copper vacancies \cite{Szaniawski2015, Witte2014}.
Finally, for the initial condition, we assume that a mixture of CuIn and CuGa binary species is an adequate simplified representation of an actual metal precursor, which typically contains more complex Cu-Ga-In alloys and elemental In \cite{Purwins2007}. 
Consequently, we propose that the system can modeled using the mechanism in Table \ref{tab:rxn_list} and the values given for propensity constants are discussed in the following section.
\begingroup
  \squeezetable
\begin{table}
  \caption{Reaction mechanism/lattice events for chalcopyrite production with values of baseline propensity constants ($p$). Propensity constants, as shown in Section \ref{sec:properties}, depend on the size of the lattice element, $\ell$. In this example, $\ell=$ 100~nm. Each reaction takes place between exactly two lattice sites; for accurate stoichiometry, Se and S sites are twice as number-dense, and chalcopyrite sites are half as number-dense as other species; \textit{atomic} number density is therefore constant across lattice sites.\label{tab:rxn_list}}
\begin{ruledtabular}
\begin{tabular}{cld}
  \multirow{1}{.65in}{Event Class} & \multicolumn{1}{c}{Event} & \multicolumn{1}{c}{$p$} \\
  \midrule
  \multirow{4}{.65in}{Adsorption/\\Desorption} & Selenium Adsorption &0.20\\
                                               & Sulfur Adsorption &0.01\footnote{H\textsubscript{2}Se-only simulations are also presented, in which case sulfur adsorption propensity is zero.}\\
                                               & Selenium Desorption &5.00\\
                                               & Sulfur Desorption &5.00\\
  \midrule
  \multirow{4}{.65in}{Binary \\Selenide/\\Sulfide Formation} & \ce{CuIn + 2Se -> CuSe + InSe} &50.00\\
                                                             &\ce{CuGa + 2Se -> CuSe + GaSe} &1.00\\
                                                             &\ce{CuIn + 2S -> CuS + InS} &1.00\\
                                                             &\ce{CuGa + 2S -> CuS + GaS} &25.00\\
  \midrule
  \multirow{8}{.65in}{Chalcopyrite Formation} &\ce{CuSe + InSe -> CuInSe2} &0.10 \\
                                              &\ce{CuSe + GaSe -> CuGaSe2} &0.10 \\
                                              &\ce{CuS + InS -> CuInS2} &0.10 \\
                                              &\ce{CuS + GaS -> CuGaS2} &0.10\\
                                              &\ce{CuS + InSe -> 0.5CuInSe2 + 0.5CuInS2} &0.10\\
                                              &\ce{CuS + GaSe -> 0.5CuGaSe2 + 0.5CuGaS2} &0.10\\
                                              &\ce{CuSe + InS -> 0.5CuInSe2 + 0.5CuInS2} &0.10\\
                                              &\ce{CuSe + GaS -> 0.5CuGaSe2 + 0.5CuGaS2} &0.10\\
  \midrule
  \multirow{9}{0.65in}{Diffusion\footnotemark[2]}\footnotetext[2]{All pairs of species not shown here have baseline diffusion propensities of zero. Refer to Section \ref{sec:reduction} for justification of zero-valued diffusion propensities.} & \ce{CuIn <-> CuGa} & 20.00 \\
                        & \ce{2Se <-> CuInSe2} & 1.00\\
                        & \ce{2Se <-> CuGaSe2} & 1.00\\
                        & \ce{2Se <-> CuInS2} & 1.00\\
                        & \ce{2Se <-> CuGaS2} & 1.00\\
                        & \ce{2S <-> CuInSe2} & 10.00\\
                        & \ce{2S <-> CuGaSe2} & 10.00\\
                        & \ce{2S <-> CuInS2} & 10.00\\
                        & \ce{2S <-> CuGaS2} & 10.00\\
\end{tabular}
\end{ruledtabular}
\end{table}
\endgroup

\subsection{Parameter Reduction}
\label{sec:reduction}
The reaction mechanism presented in Table \ref{tab:rxn_list} involves 12 reaction propensities, 4 adsorption/desorption propensities, and, in principle, ${14 \choose 2} = 91$ unique diffusion propensities (though many are neglected).
Furthermore, the lattice size may affect the model results (see Section \ref{sec:properties})  and will greatly affect the computation time.
With such a large parameter set and a computationally intensive model, conventional parameter fitting is impractical.
We present three simplifying assumptions and heuristics and show how we can use them to guide us in determining physically meaningful estimates for the model parameters.
\begin{enumerate}
  \item\textbf{Parabolic Film Growth:} 
Results from the literature \cite{Purwins2006, Kim2005} suggest that Cu(InGa)(SeS)\textsubscript{2} films produced via reaction of metal precursors follow a parabolic growth mechanism, referring to a solid state reaction process where there is a planar, advancing reaction front, rather than a nucleation and growth mechanism.
Invoking this mechanism suggests that the gas phase reactants, Se and S, can diffuse through reacted species, but not through the original CuIn and CuGa species.
Therefore, diffusion propensities are set such that no species can diffuse with precursors---except for precursors themselves (CuIn and CuGa), which may diffuse with each other.

Recognizing that the diffusion of Se and S are rate limiting in the parabolic mechanism, we can estimate the magnitude of the diffusion coefficients using the characteristic diffusion time:
\begin{equation}
  \label{eqn:diff_time}
  \tau_D=\frac{L^2}{D_{s_1,s_2}}
\end{equation}
Since the reaction takes place on the order of minutes and the film thickness is approximately $2~\mu m$, a reasonable estimate for the diffusivity of Se is $6.7 \times 10^{-14} m^{2}s^{-1}$.
After selecting a lattice size, the diffusion coefficients can be used to estimate the propensity constant for diffusion from Equation \ref{eqn_diffusivity}.
In the simulations presented below, the lattice size is 100 nm, suggesting 1.7 (which we truncate to 1.0) is a reasonable estimate of selenium diffusion propensity.

\item\textbf{Slow Chalcopyrite Interdiffusion:} 
One feature commonly observed in Cu(InGa)(SeS)\textsubscript{2} films produced by reaction of metal precursors is a persistent gradient in gallium.
That is, a gallium gradient forms and does not quickly anneal out to yield a uniform film \cite{Jensen1993,Witte2014,Kim2012,Hergert2006,Marudachalam1997}.
However, because  Cu(InGa)(SeS)\textsubscript{2} is a continuous solid solution, the gradient is not a thermodynamic limitation, but must be limited by mass transfer.
Therefore, we set the diffusion propensities of fully reacted species with each other to be zero, as their diffusion time is longer than the time scale of a typical reaction.

\item\textbf{Fast Precursor Interdiffusion:} 
In contrast to fully reacted chalcopyrite species, the unreacted species must interdiffuse relatively quickly.
The fast interdiffusion of CuIn and CuGa (at least faster than the time scale of the diffusion and reaction of Se) is required for the reaction asymmetry to cause composition gradients.
\end{enumerate}

The three heuristics presented above were used to guide parameter selection, especially the diffusion propensities for all sulfur-free lattice events (sulfur-containing lattice events are discussed in the next section when sulfur-containing models are presented).
From the first heuristic, to ensure a reasonable time scale, the diffusion propensities of selenium with chalcogenides and chalcopyrites were set to 1.0.
Based on the second heuristic, most of the remaining diffusion propensities were set to zero, except for interdiffusion of precursor species (CuIn and CuGa).
The third heuristic compels us to select a diffusion propensity for CuIn and CuGa that is larger than that of selenium and reacted phases; therefore we selected 20.0 for the baseline.
As discussed earlier, the rate of reaction of Se and In species is much faster than Se and Ga, therefore a baseline estimate for the propensity constant is 50.0 for InSe formation and 1.0 for GaSe formation.
The remaining reaction propensities (chalcopyrite formation reactions) did not have a substantial effect on the composition profile and were set to 0.1.
One possible approach for estimating adsorption and desorption propensities would be to use first principles and the equations derived in Section \ref{sec:properties}.
However, unless reasonable estimates for sticking coefficient are available (in many cases, its order of magnitude is much less than 1), we recommend choosing a value of similar order to the propensity constants of the other processes to reduce computational burden.
For our case, because we do not expect significant accumulation of selenium \textit{in its elemental form} (due to its high vapor pressure and the stability of H\textsubscript{2}Se molecules), the propensity of adsorption should be less than propensity for desorption.
Therefore, for selenium, we set adsorption propensity to 0.1 and desorption propensity to 5.0.
These propensities were used for the baseline simulations presented in the next section and are shown in Table \ref{tab:rxn_list}.

\subsection{Composition Profile Prediction}
\begin{figure}
    \centering
    \includegraphics[width=3.5in]{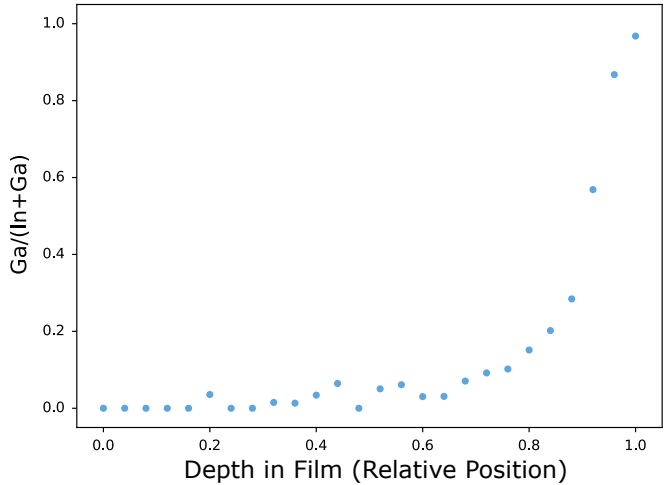}
    \caption{Ga  profile produced from simulation of an H\textsubscript{2}Se-only process. Note that ``relative position" is a scaled variable. The full thickness is approximately 2 $\micro m$.}
    \label{fig:Se_only}
\end{figure}
\begin{figure}
    \centering
    \includegraphics[width=3.5in]{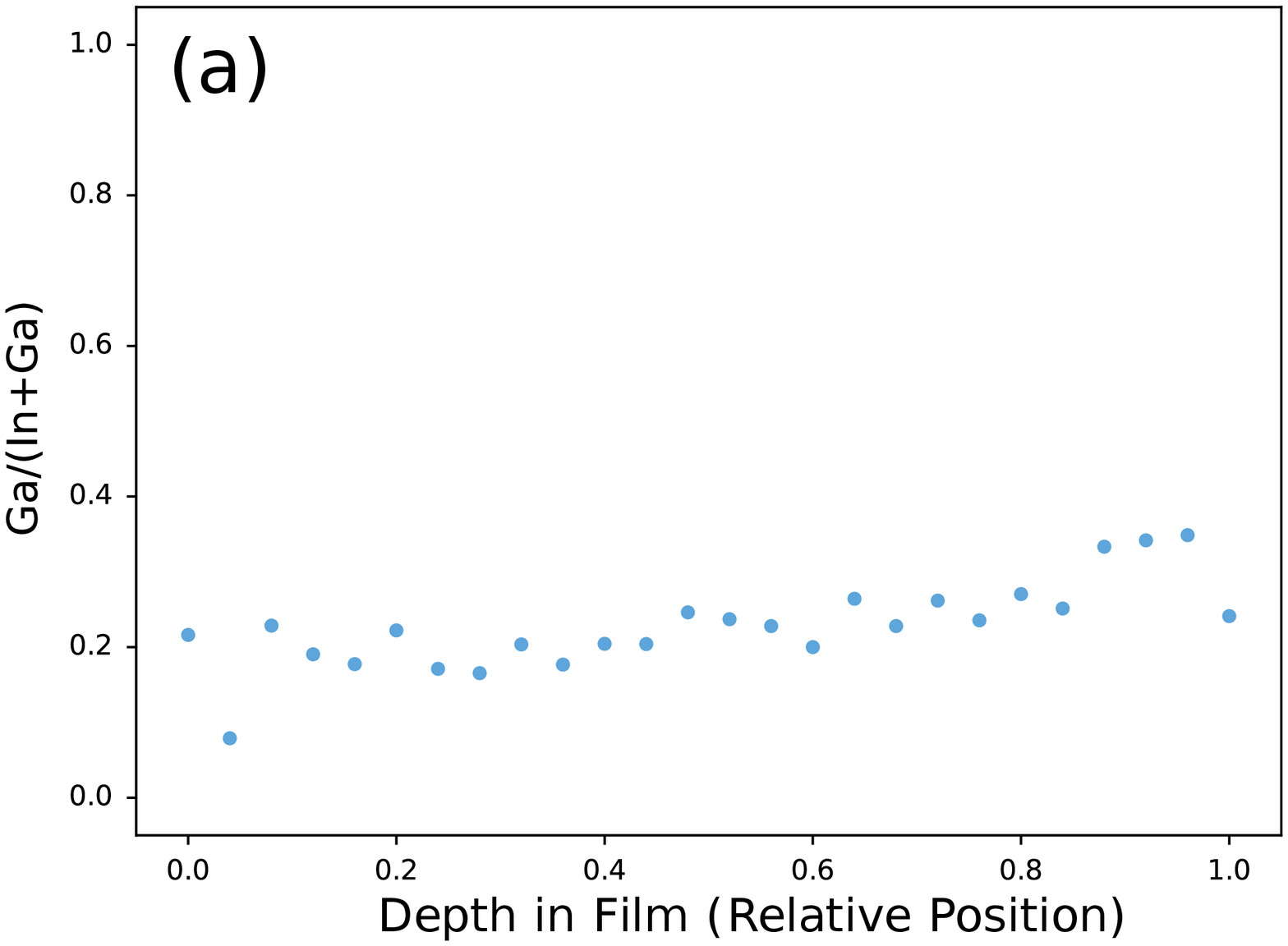}
    \includegraphics[width=3.5in]{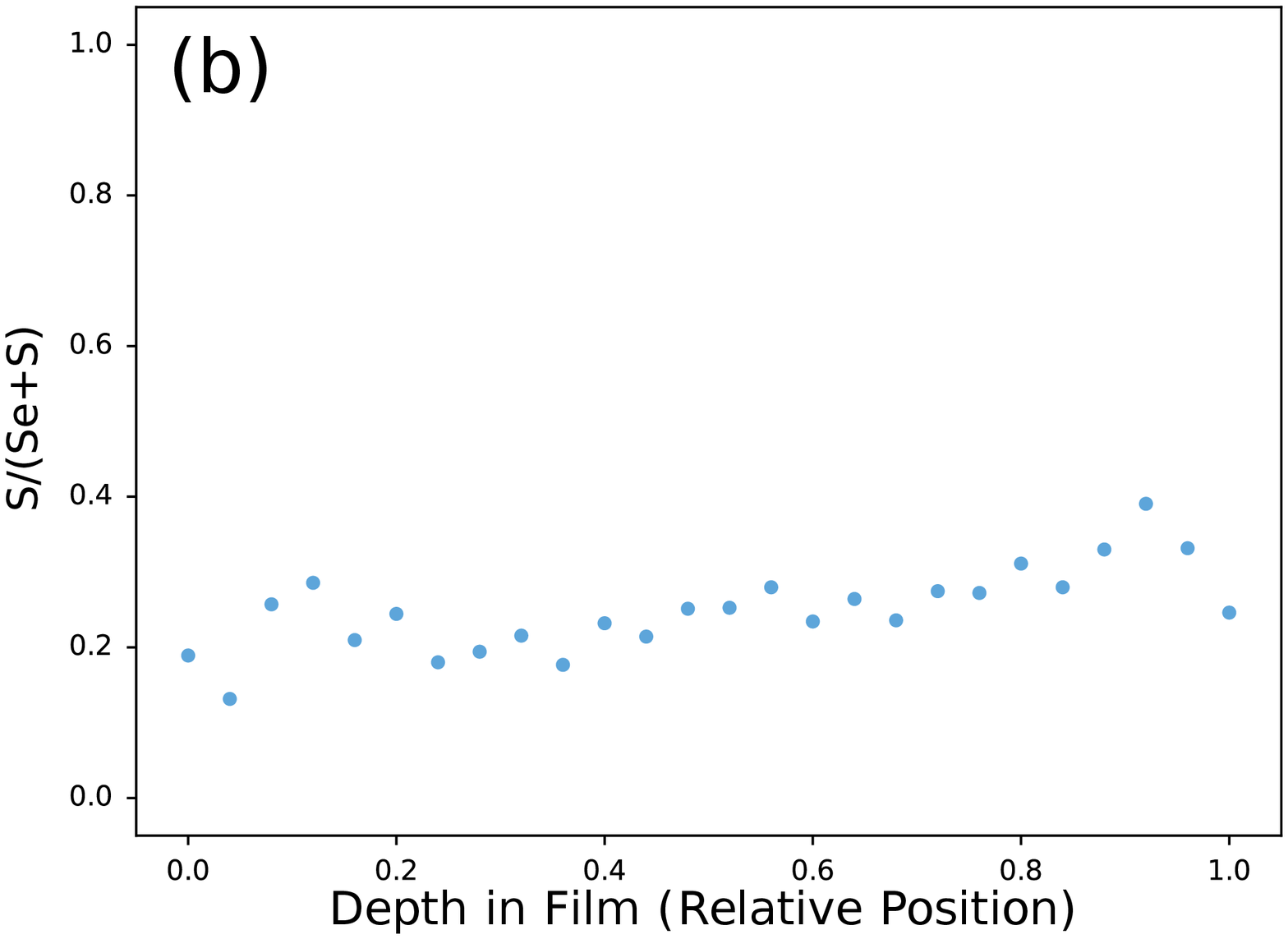}
    \caption{Profiles of Ga (a) and S (b) produced from simulation of an H\textsubscript{2}Se+H\textsubscript{2}S process.}
    \label{fig:simultaneous}
\end{figure}

The simulation algorithm from Section \ref{sec:approach} was applied to the chalcopyrite growth model in Table \ref{tab:rxn_list}.
The initial lattice, $\mathbf{L_0}$, is 23 rows $\times$ 100 columns, with rows 0 to 11 specified as vacancy elements, and rows 12 to 22 specified as a 0.25:0.75 mixture of CuGa and CuIn elements (this ratio was chosen for industrial relevance).
Because Cu(InGa)(SeS)\textsubscript{2} films for photovoltaic cells are typically about 2 $\micro m$ thick, the 23 rows result in lattice length $\ell\approx100$ nm.
The algorithm was applied to advance the lattice until 40 minutes simulation time elapsed.
A useful way to visualize the results from the simulations is to plot the Ga/(In+Ga) and S/(S+Se) ratios as functions of depth.

For a first case, we considered a process with only H\textsubscript{2}Se, for which the propensity for adsorption of H\textsubscript{2}S was therefore set to zero.
The results are shown in Fig.\@ \ref{fig:Se_only}, where a large amount of gallium accumulates near the back of the film, which is a feature that has been well-documented experimentally \cite{Jensen1993, Kim2012, Hanket2007}.

For a second case, the precursors are reacted simultaneously in equal concentrations of H\textsubscript{2}Se and H\textsubscript{2}S.
The simultaneous reaction process is modeled by setting adsorption of sulfur lower than adsorption of selenium, but diffusion of sulfur faster than diffusion of selenium (the effect of sulfur adsorption is discussed in more detail in Section \ref{sec:valid}).
Through-film profiles of gallium and sulfur are shown in Fig.\@ \ref{fig:simultaneous}.
Consistent with experimental results from \cite{Kim2011}, gallium is distributed more homogeneously than in a process with H\textsubscript{2}Se only.

\begin{figure}
    \includegraphics[width=3.5in]{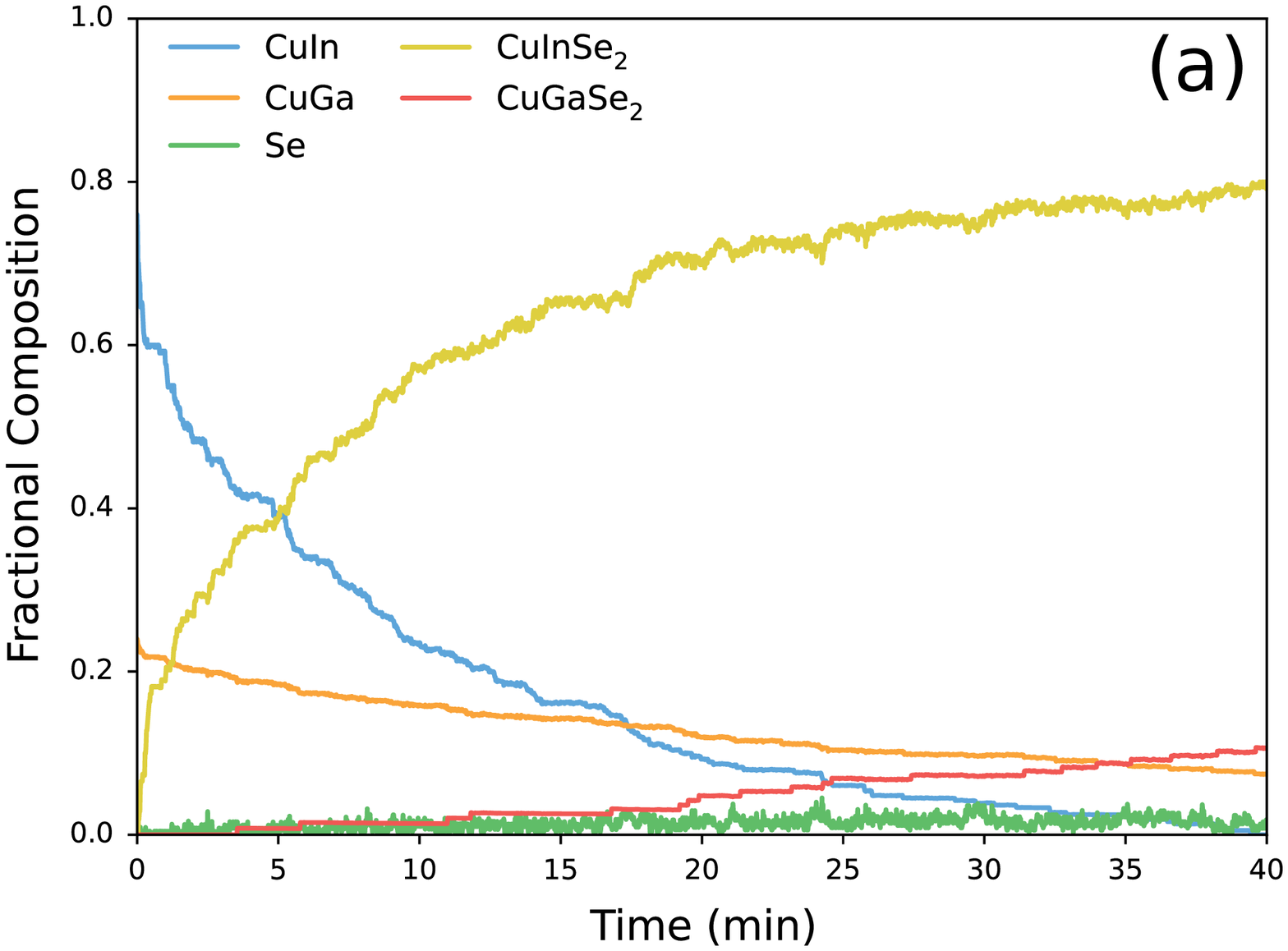}
    \includegraphics[width=3.5in]{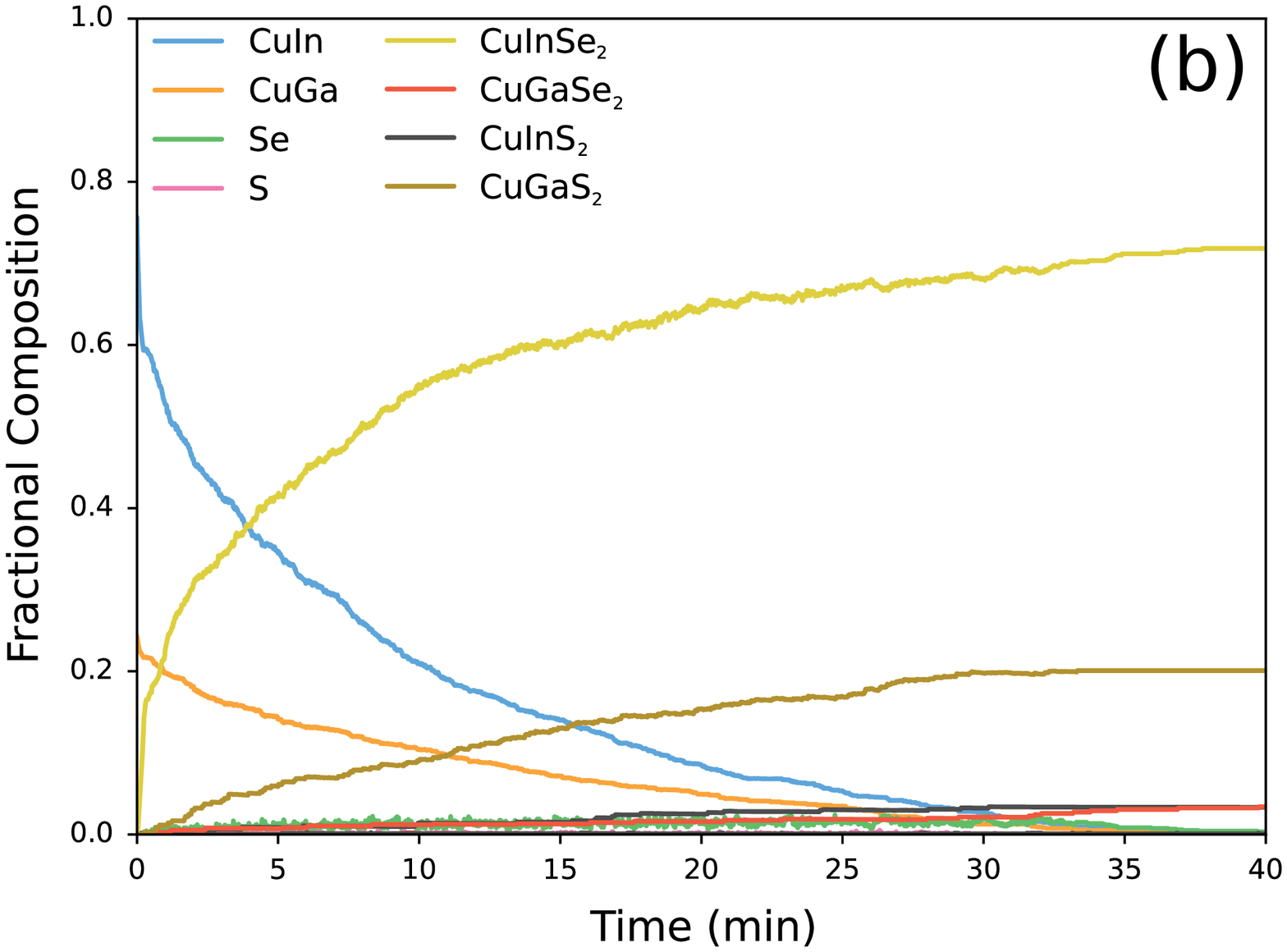}
    \caption{Time evolution of composition of the reacting film for the reaction in (a) H\textsubscript{2}Se only and (b) H\textsubscript{2}Se and H\textsubscript{2}S. Fractional composition is the fraction of occupied lattice sites, or volume fraction of the film. The intermediate phases, binary selenides and sulfides, are not shown but exist as a small volume fraction throughout the simulation.}
    \label{fig:comp_evo}
\end{figure}

The gallium and sulfur profiles in Figs.\@ \ref{fig:Se_only} and \ref{fig:simultaneous} result from complex interactions of the different propensity constants.
One way to understand how these profiles arise is to examine the dynamics of the process.
A simplified representation of the dynamics is shown in Fig.\@ \ref{fig:comp_evo}, where composition is spatially averaged through the entire film and shown as a function of time.
In the first case, because the propensity for reaction with CuIn is much faster than with CuGa, most CuGa remains in its precursor form until nearly all of the CuIn is converted to chalcopyrite.
During this time, CuGa diffuses toward the back contact leading to the profile in Fig.\@ \ref{fig:Se_only}.
The observation that Ga-containing species diffuse toward the back contact, while In-containing species are incorporated in the chalcopyrite phase (CuInSe\textsubscript{2}), is supported by experiments from the literature.
For example, in \cite{Hanket2007}, x-ray diffraction patterns indicate that Cu\textsubscript{9}(In\textsubscript{.2}Ga\textsubscript{.8})\textsubscript{4} is present on the back contact after a 10 minute reaction in H\textsubscript{2}Se, but only Cu\textsubscript{9}Ga\textsubscript{4} is present after 30 and 90 minute reactions.
For the second case, with H\textsubscript{2}Se and H\textsubscript{2}S, there is first a rapid increase in CuInSe\textsubscript{2} because of the relatively faster adsorption of Se compared to S; however, the faster diffusion of sulfur, combined with its preference to react with CuGa, leads to more incorporation of CuGa earlier in the process than if there were H\textsubscript{2}Se alone.
\subsection{Agglomeration Size Distribution}
\label{sec:size_dist}
One of the advantages of applying stochastic simulation at the mesoscopic scale is that it allows one to study fluctuations in local composition explicitly.
The fluctuations result in some spatially confined features that can be described statistically in the example system, Cu(InGa)(SeS)\textsubscript{2}. 
One of the important features observed experimentally is agglomerations of Ga-rich species at the back contact \cite{Hanket2007}.
In principle, there are two causes of these agglomerations: thermodynamic phase separation, and random fluctuations in species positions.
The modeling approach cannot account for thermodynamic phase separations, but we can use it to examine the effect of random fluctuations.
Although thermodynamic driving forces are likely important in Cu(InGa)(SeS)\textsubscript{2} films (evidenced by formation of Cu\textsubscript{9}Ga\textsubscript{4} phase near the back contact \cite{Hanket2007, Kim2012}), we still present the model results to demonstrate our approach for characterizing agglomeration size distribution.

\begin{figure*}
    \includegraphics[width=2.5in]{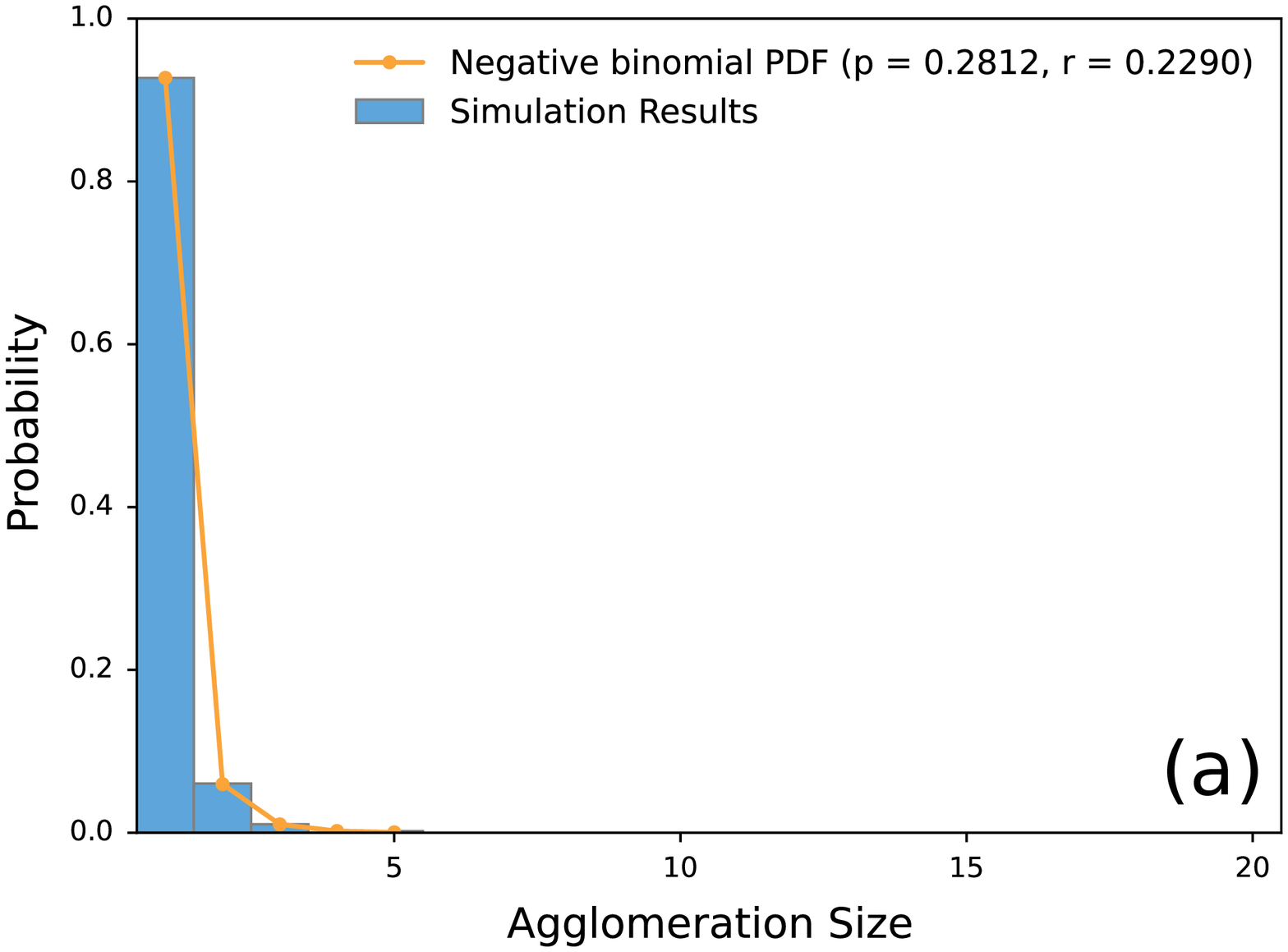}
    \includegraphics[width=2.5in]{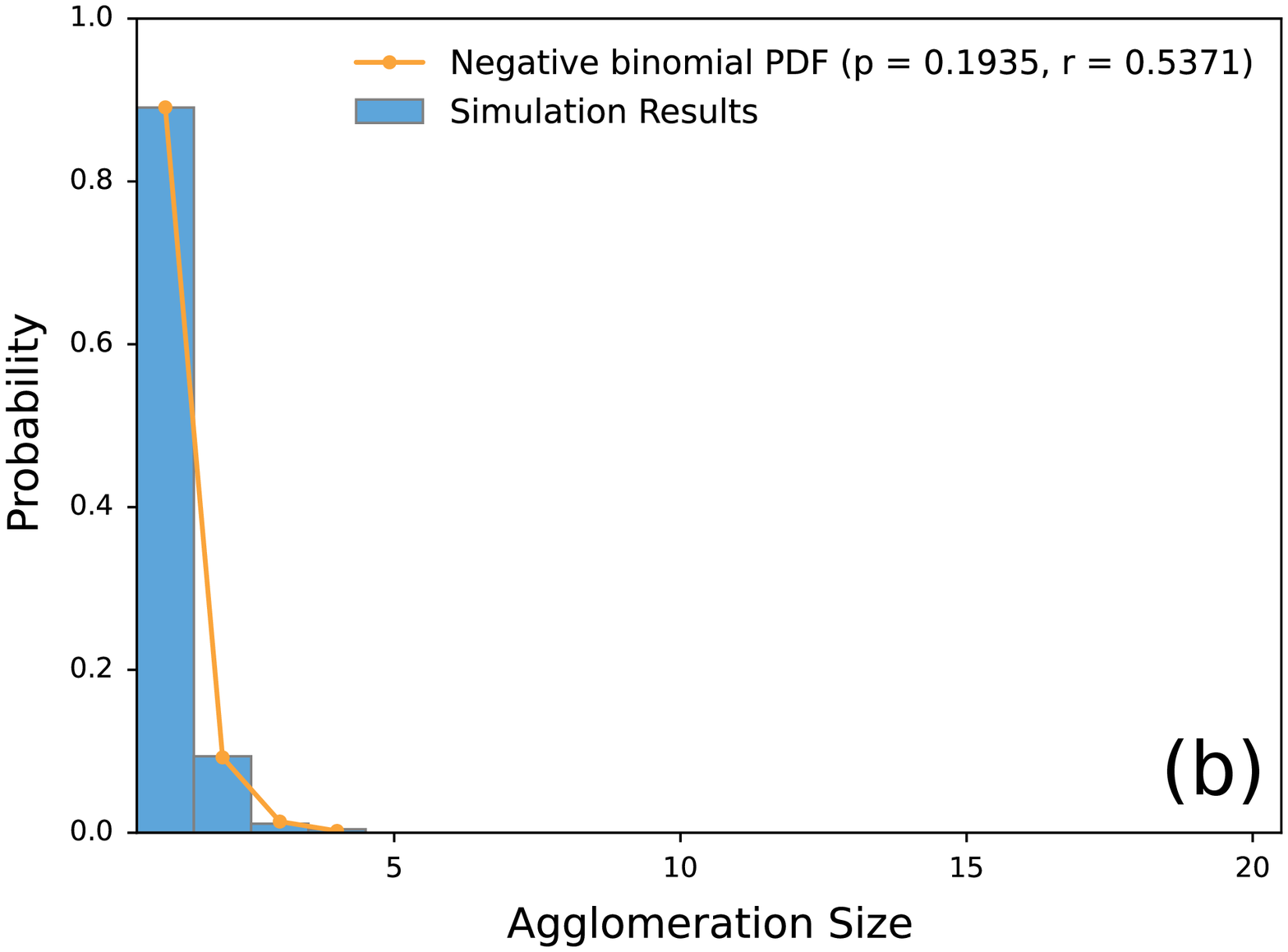}
    \includegraphics[width=2.5in]{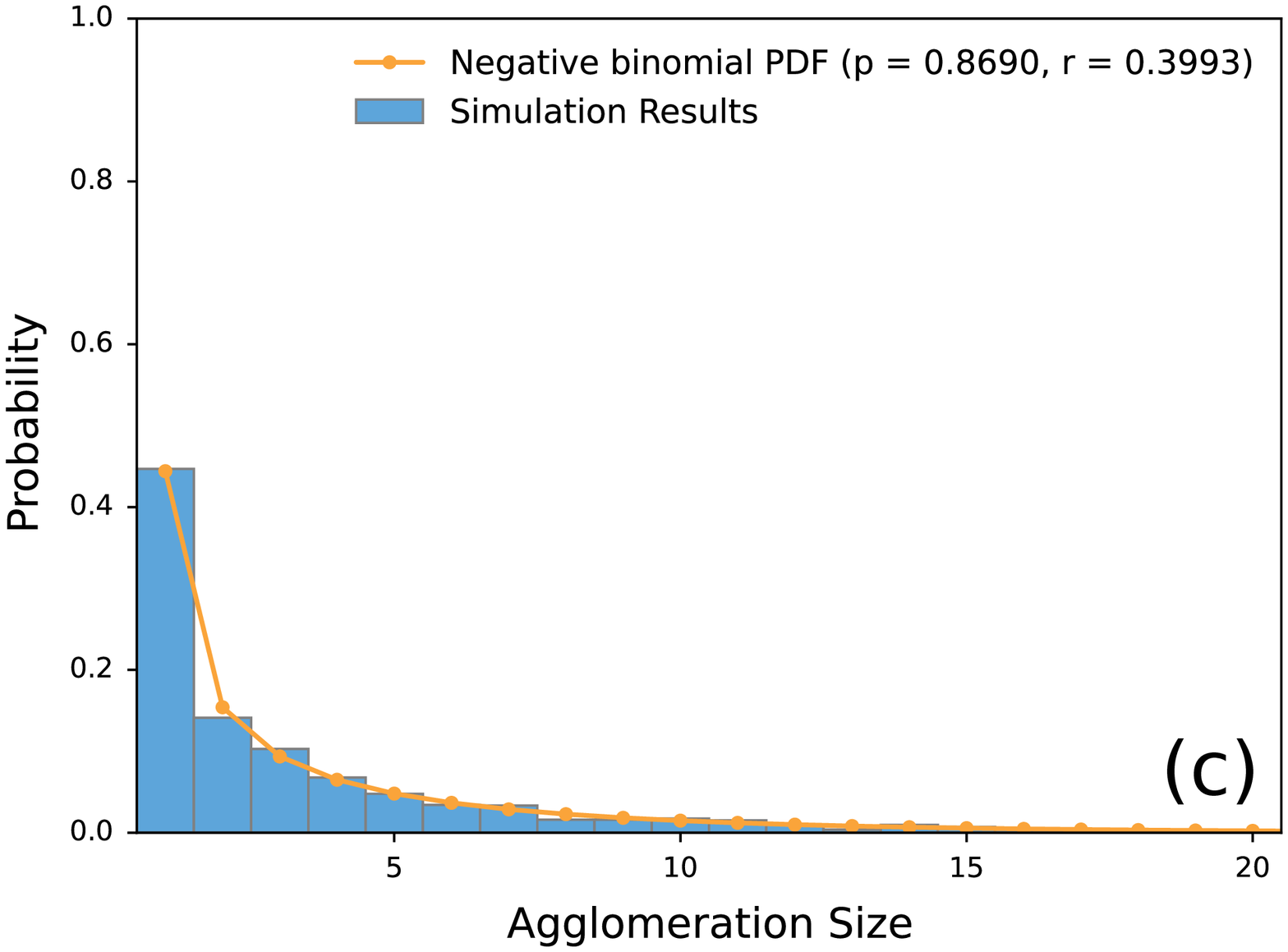}
    \includegraphics[width=2.5in]{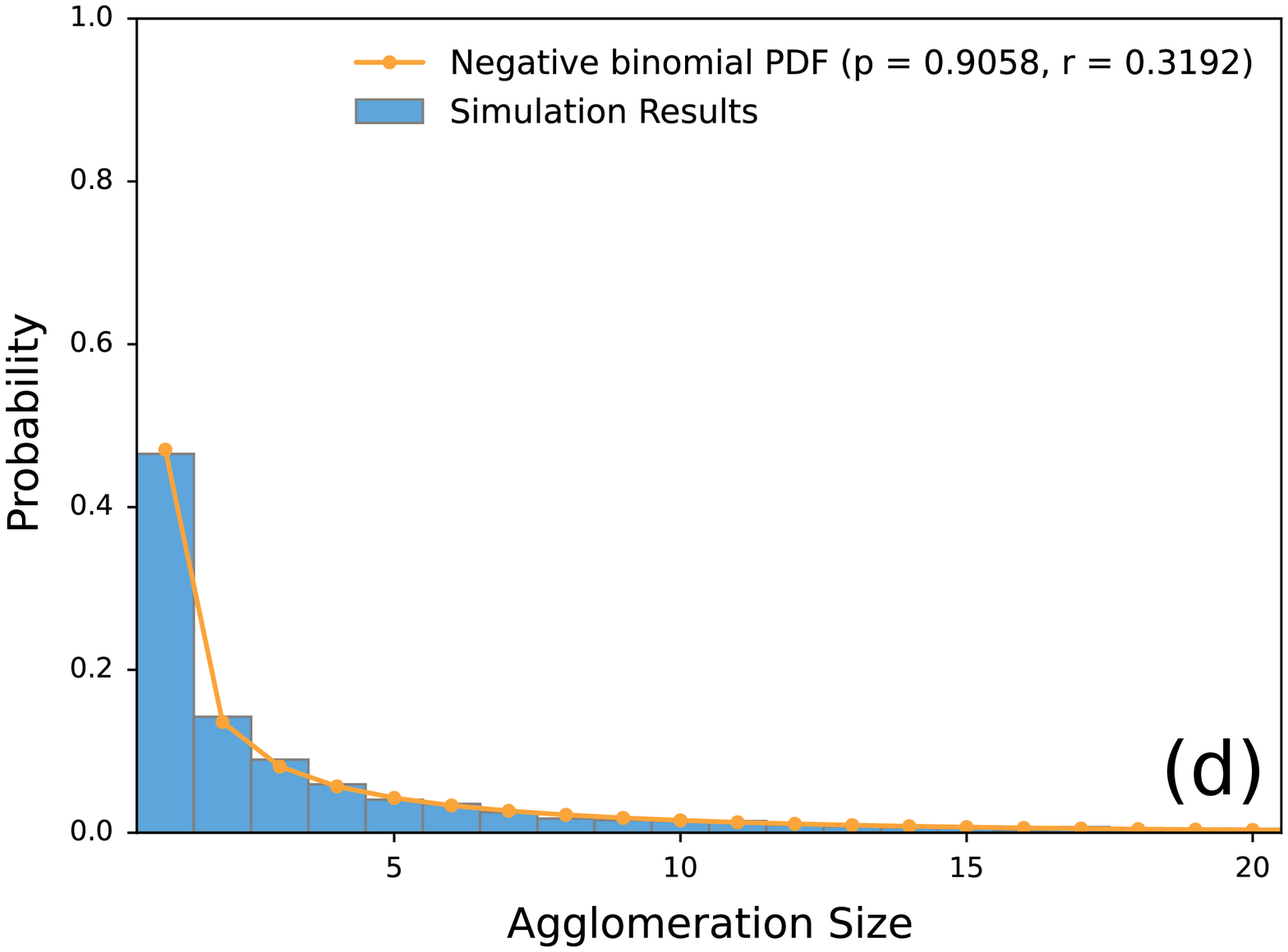}
    \includegraphics[width=2.5in]{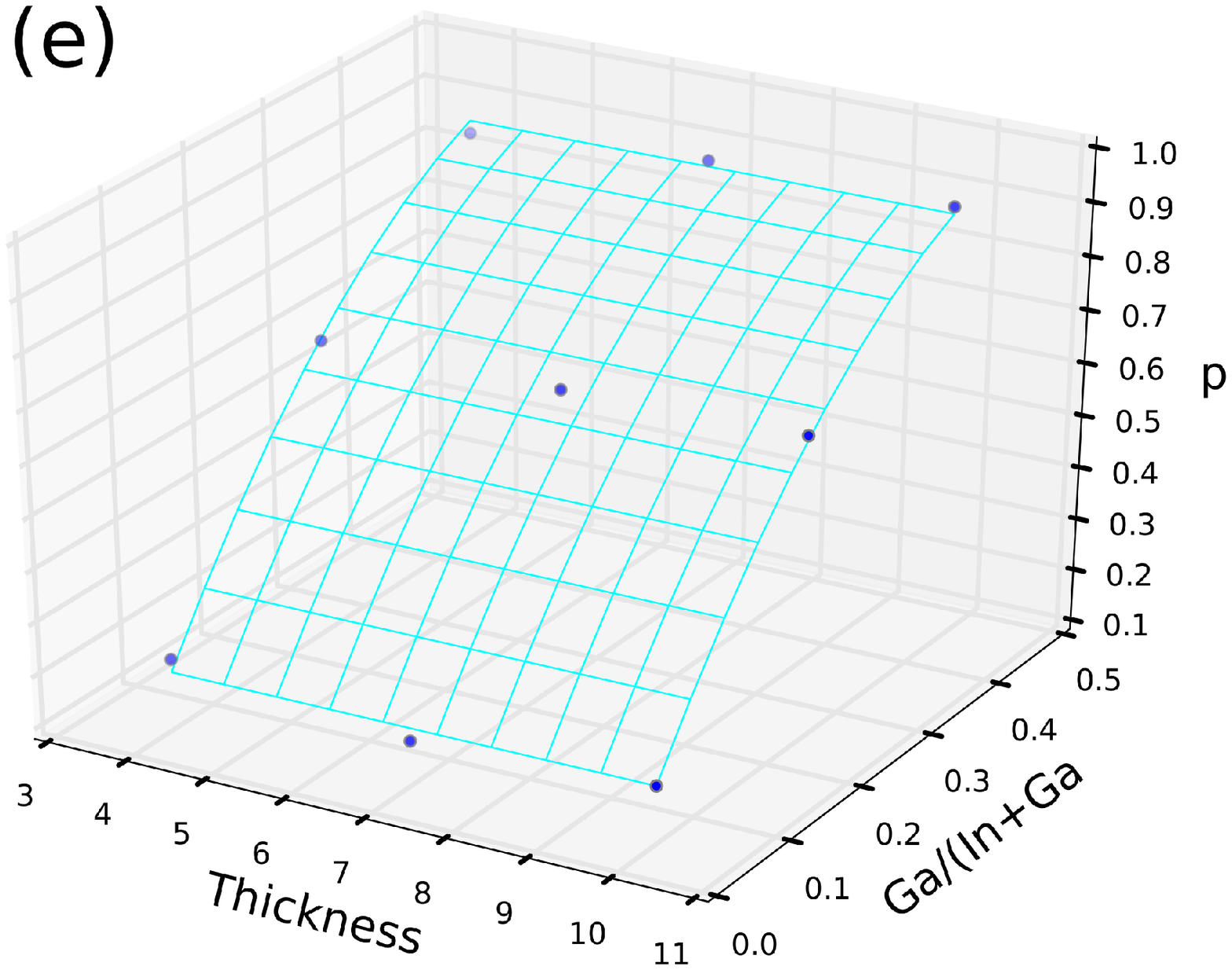}
    \includegraphics[width=2.5in]{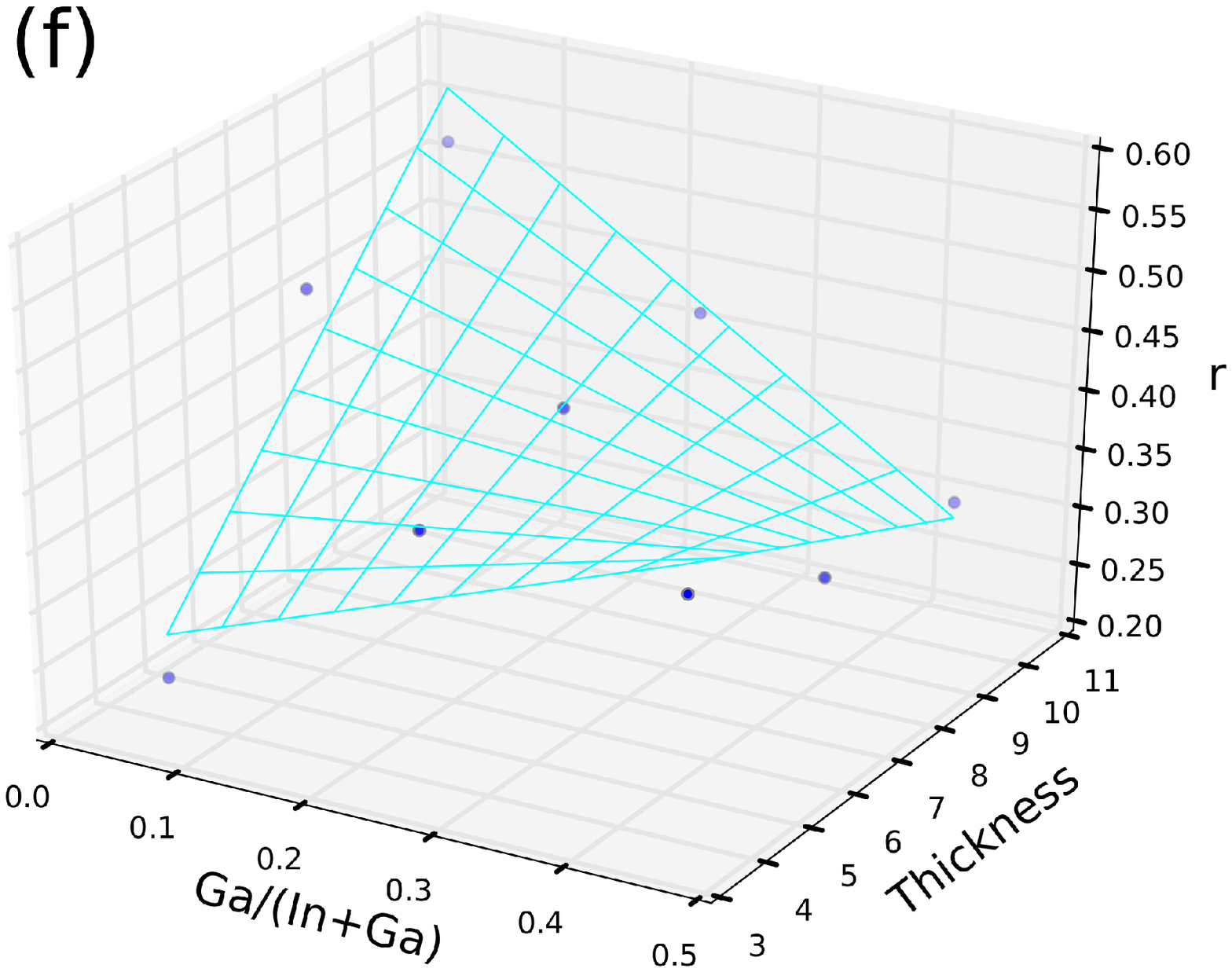}
    \caption{(a)--(d) Agglomeration size distribution for CuGa after 10 s simulation time with no reactions/adsorption allowed, for the following compositions (Ga/(In+Ga)) and lattice thicknesses (in units of lattice sites), respectively: (a) 0.05, 4; (b) 0.05, 10; (c) 0.45, 4; (d) 0.45, 10. (e) MLE estimates (points) and fitted response surface for negative binomial random variable parameter $p$. (f) MLE estimates (points) and fitted response surface for negative binomial random variable $r$.}
    \label{fig:agglom}
\end{figure*}

For this purpose, simulations with large lattice sizes (2500 columns) were run with no adsorption or reaction events allowed and with precursors (\textit{i.e.}, $\mathbf{L_0}$ matrices) of varying composition and thickness.
We define \textit{any} isolated collection of CuGa elements to be an agglomeration, regardless of size.
Thus, a single, isolated CuGa element is considered a size-1 agglomeration.
Figs.\@ \ref{fig:agglom}a--d show histograms of the size distribution of agglomerations for different film thicknesses and compositions (which models later stages of the reaction, where the precursor film gets thinner as the front of the film is converted to chalcopyrite).

We consider the negative binomial probability distribution function as an appropriate model for quantifying the effects of composition fluctuations on agglomeration formation for the following reasons:
The negative binomial random variable is, by definition, the number of Bernoulli trials with probability $p$ until $r$ ``successes" are reached.
Further, $r$ can be thought of as a ``waiting time" parameter and need not be integer-valued.
More generally, the negative binomial random variable is an overdispersed form of a Poisson random variable with mean $\mu = \frac{pr}{1-p}+1$, but with gamma-distributed intensity.
In this form, the $r$ parameter is often referred to as an ``aggregation" or inverse dispersion parameter, where as $r \rightarrow 0$, $\sigma^2 \rightarrow \infty$, where $\sigma^2$ is variance \cite{Lloyd-Smith2007, Shaw1998, Alexander2000}.
Now, if the formation of an agglomeration can be considered as a series of trials, where $p$ is the probability of adding another species to the agglomeration, then the agglomeration size should follow a negative binomial distribution with $r=1$ (\textit{i.e.}, a geometric distribution).
However, because of its improved handling of highly dispersed data, we use the more general negative binomial form allowing both $p$ and $r$ to be estimated.
Intuitively, an increasing $p$ or increasing $r$, will move the distribution rightward, toward a tendency for larger agglomerations.

The expression for the negative binomial probability distribution,
\begin{equation}
  \label{eqn:nb}
  f(x|r,p)=\frac{\Gamma(x-1+r)}{(x-1)!\Gamma(r)}p^x(1-p)^r, \mathrm{~for~} x = 1,2,3...
\end{equation}
is used to characterize agglomeration size distribution by estimating the two parameters, $p$ and $r$, using Maximum Likelihood Estimation (MLE) (see Appendix \ref{app:MLE}) from the simulation data.

In order to uncover the effects of geometry (\textit{i.e.}, film thickness) and composition (Ga/(In+Ga)) on the parameters, $(p,r)$, we constructed a face-centered cubic response surface design, and used it to develop an empirical model.
In the response surface model, geometry and composition are considered factors that affect the response variables, $p$ and $r$.
Nine simulations were run with varying lattice thickness and Ga/(In+Ga) ratios; $p$ and $r$ were estimated with MLE; and the estimates were used to fit 2\textsuperscript{nd} order polynomial models of $p$ and $r$ as a function of lattice thickness and composition (see Appendix \ref{app:RSP}).
The fitted response surfaces, \textit{i.e.}, the polynomial models, are shown in Figs. \ref{fig:agglom}e--f, from which we conclude that the probability of success parameter, $p$, is a function of only composition.
Thus, all else being equal, when the fraction of CuGa sites increases, the agglomerations of CuGa tend to grow larger.
The waiting time/aggregation parameter, $r$, has a more complicated response surface that is strongly affected by the interaction of geometry and composition.
Specifically, from the partial derivative of the response surface ($\frac{\partial r}{\partial x_1}$ of Equation \ref{eqn:rsp_fit}), films with a gallium fraction less than 0.33 form larger agglomerations (due to larger $r$) when thickness is greater; the inverse is true for films with a gallium fraction greater than 0.33.

\subsection{Model Predictions vs.\@ Experimental Results}
\label{sec:valid}

\begin{figure}
  \includegraphics[width=3.5in]{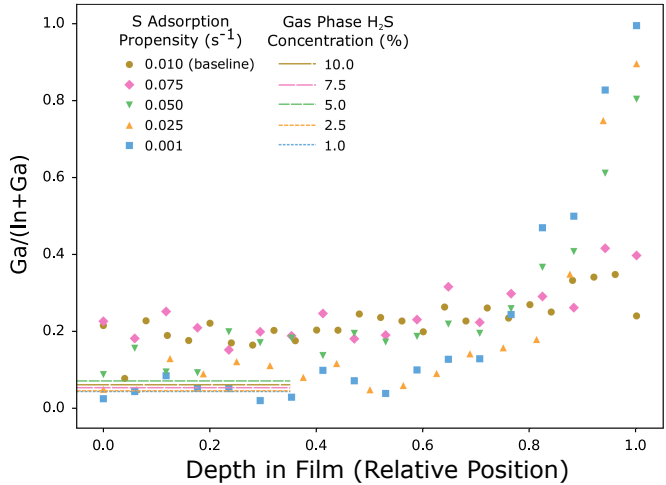}
  \caption{Gallium profiles (points) resulting from simulation of an H\textsubscript{2}Se+H\textsubscript{2}S process using baseline simulation parameters (from Table \ref{tab:rxn_list}) except for sulfur adsorption, which has a varying propensity constant. EDX measurements (horizontal lines, with length corresponding roughly to sampling depth) of Ga/(In+Ga) from samples produced with varying H\textsubscript{2}S concentration.}
  \label{fig:vary_S}
\end{figure}

In order to compare our model predictions to experimental data, we produced a series of samples in the laboratory with the following process conditions: reaction temperature of 550~\degree C, H\textsubscript{2}Se concentration of 1\%, and varying H\textsubscript{2}S concentration in Ar gas.
(See Appendix \ref{app:exp} for a summary of experimental methods).
Sample compositions were measured using Energy Dispersive X-ray (EDX) spectroscopy with results for Ga/(In+Ga) shown in Fig.\@ \ref{fig:vary_S}.
While sulfur was detected in each sample, the composition was too low (less than 1 atom percent) for accurate quantitative measurement.
The EDX spectroscopy measurements have a depth sensitivity of approximately 0.7 to 0.9 $\micro m$ (or about one half of total film thickness).
Therefore, EDX measurements can be treated as spatial averages of composition in the front half of the film.
Each precursor sample has an average Ga ratio of 0.25; therefore, if the measured Ga ratio is less than 0.25, then the  Ga is segregated toward the back of the film.

Without time- and depth-resolved data, conventional parameter fitting is impossible.
Further, even if the data were available, parameter fitting would be very computationally intensive because it requires numerous simulation runs.
Instead, we compare the effect of varying H\textsubscript{2}S concentration in the gas phase to the equivalent change in our model, that is, changing the propensity for adsorption of sulfur.
We expect that varying H\textsubscript{2}S concentration will affect the through-film gallium profile, and that this effect will be captured in experimental results and by our model.
However, rather than absolute convergence (which would require more precise parameter estimates), similar trends should be observed in experiments and simulations.

Fig.\@ \ref{fig:vary_S} shows the through-film gallium profiles from the model using the baseline simulation parameters, except for adsorption of sulfur, which varies between simulations.
Also displayed in the figure is the measured gallium ratio of the films produced with varying H\textsubscript{2}S concentration in the gas phase.
In both simulation and experiment, when the sulfur (propensity in simulations or concentration in experiments) is low, gallium fraction increases near the front surface with increasing sulfur, but the effect is diminished as sulfur increases further.
However, our simulation overestimates the effect of sulfur on gallium fraction.
Several mechanisms may explain the discrepancy; for example, the reaction between selenium and indium or the adsorption rate of selenium may be faster than our estimate, which limits the tendency of sulfur to increase gallium homogenization.
Other authors \cite{Kim2011} reported more substantial gallium homogenization than we observed, which may be explained by the lower H\textsubscript{2}Se concentration (0.35\%) in their experiments.
However, we cannot use such low concentrations in our reactor, which operates in batch mode, because the gas phase would be depleted of H\textsubscript{2}Se before reaction is complete (the authors of \cite{Kim2011} use a flow reactor, where H\textsubscript{2}Se is continuously replenished).

\section{Summary and Conclusions}

We have presented a novel method for modeling thin film growth using a stochastic simulation method.
We demonstrated an algorithm that makes the model computationally tractable for a typical desktop PC.
In particular, we showed how the model applies to the industrially relevant case of thin film Cu(InGa)(SeS)\textsubscript{2} growth using a precursor reaction process.
The model explains how the complicated, experimentally observed through-film profiles in Ga and S arise from complex interactions of reaction rates, adsorption rates, and diffusion limitations.
We show that the stochastic nature of the model allows it to be used to understand lateral inhomogeneities, such as agglomerations, that would otherwise be ignored or averaged out in continuum approaches.

We believe that this modeling approach can find wide application in a number of thin film or other solid state material systems.
In particular, we suggest that this approach will be especially useful for vacuum-deposited Cu(InGa)Se\textsubscript{2}, for Cu\textsubscript{2}ZnSnS\textsubscript{4}, and for silicon systems with impurities because of similarities to our model system, including the potential for composition profiles and lateral heterogeneity.
Furthermore, because of our method's emphasis on material adjacency, it can be applied to systems with complex geometry, such as graphene or carbon nanotubes. 
More fundamentally, however, the stochastic simulation that we developed is a new approach that could be generalized for any system where system evolution is governed by network connectivity (in our case, material adjacency) instead of bulk composition, and may be applied in entirely unrelated fields.

\section*{Acknowledgment}
This material is based upon work primarily supported by the Quantum Energy and Sustainable Solar Technology Engineering Research Center (QESST ERC), the National Science Foundation (NSF) and the Department of Energy (DOE) under NSF CA No. EEC-1041895.  Any opinions, findings and conclusions or recommendations expressed in this material are those of the author(s) and do not necessarily reflect those of NSF or DOE.
\appendix
\section{Parameter Estimation and Validation for Agglomeration Distribution}
\label{app:MLE}

In this appendix, we use a slightly altered version of the negative binomial probability model (first shown in Equation \ref{eqn:nb}), given by 
\begin{equation}
  \label{eqn:nb_app}
  f(x|r,p)=\frac{\Gamma(x+r)}{(x)!\Gamma(r)}p^x(1-p)^r, \mathrm{~for~} x = 0,1,2...
\end{equation}
to characterize the agglomeration size distribution.
In this form, the domain of $x$ has been shifted leftward by 1 unit; thus the data should be shifted down by 1 unit (\textit{i.e.}, size-1 agglomerations will be considered size-0 agglomerations), but the parameter values remain unchanged.
The two parameters, $r$ and $p$ were estimated using Maximum Likelihood Estimation (MLE).
The likelihood function for a sample of agglomeration data, $\mathbf{X}$ is then obtained as:
\begin{equation}
  \label{eqn:Lnb}
  L(r,p|\mathbf{X})=\prod_{x\in\mathbf{X}} \frac{\Gamma(x+r)}{(x)!\Gamma(r)}p^x(1-p)^r
\end{equation}
and for a total of $N$ agglomeration samples, the log-likelihood function is:
\begin{equation}
  \label{eqn:llnb}
  \begin{split}
    l(r,p|\mathbf{X})= &Nr\ln(1-p)-N\ln(\Gamma(r))  + \\ &\sum_{x\in\mathbf{X}}(\ln(\Gamma(x-1+r))-\ln((x-1)!)+x\ln(p))
  \end{split}
\end{equation}
To maximize the log-likelihood function (equivalent to maximizing the likelihood function) the partial derivatives of the log-likelihood function with respect to the paramters are set to zero:
\begin{equation}
  \label{eqn:dllnb}
  \begin{split}
    \frac{\partial l(r,p)}{\partial r} &= 0 = N \ln(1-p) - N \Psi(r) + \sum_{x\in\mathbf{X}}\Psi(x-1+r) \\
    \frac{\partial l(r,p)}{\partial p} &= 0 = Nrp - \frac{\sum_{x\in\mathbf{X}} x}{p}
  \end{split}
\end{equation}
where $\Psi(x)$ is the digamma function, or $\Gamma'(x)/\Gamma(x)$.  
While the second of these equations can be solved explicitly for $p$, in general, there is no closed form solution to Equations \ref{eqn:dllnb} for $r$ and $p$.
The system of equations is solved numerically to obtain the parameter estimates

\begin{table}
  \label{tab:chi}
  \caption{Results from MLE estimates and $\chi^2$ tests of the negative binomial parameters from the agglomeration distribution data. There is no evidence to reject the null hypothesis that the model is adequate in any case. The thickness and composition values were selected to complete a face-centered cubic response surface design.}
\begin{ruledtabular}
  \begin{tabular}{cccccc}
  Thickness & Ga/(In+Ga) & $N$ & $p$ & $r$ & $Pr(C^2 > \chi^2(m-3))$ \\
  \midrule
  4& .05 & 481&0.2182 & 0.2290 & N/A\footnote{Sample size too small for chi-square test. Fit is very good visually.}\\
  10& .05 & 1173&0.1935 & 0.5371 & 0.098\\
  4& .45 & 1228&0.8690 & 0.3993 & 0.526\\
  10& .45 & 2662&0.9058 & 0.3192 & 0.592\\
  4 & .25 & 1442&0.6433 & 0.4028 & 0.508\\
  10& .25 & 3454&0.6568 & 0.4352 & 0.557\\
  7 & .05 & 808 &0.1685 & 0.4826 & N/A\textsuperscript{a}\\
  7 & .45 & 1858&0.9035 & 0.3333 & 0.567\\
  7& .25 & 2428&0.6456 & 0.4280 & 0.407\\
\end{tabular}
\end{ruledtabular}
\end{table}

In order to validate this approach, simulations were run with varying film thicknesses and gallium fractions, $p$ and $r$ were estimated, and a $\chi^2$ test was applied to compare model predictions to data.
The response surface method (see Appendix \ref{app:RSP}) was used to select the specific values of film thickness and gallium fraction.
The $\chi^2$ test statistic is:
\begin{equation}
  \label{eqn:chi_sq}
  C^{2} = \sum_i^m \frac{(f_i - \phi_i)^2}{\phi_i}
\end{equation}
$f_i$ is the observed count of agglomerations of size $i$, $\phi_i$ is the predicted count using the negative binomial distribution with MLE parameter estimates, and $m$ is the largest agglomeration size with at least 5 instances.
If the model is appropriate, $C^2$ should approximate a $\chi^2(\nu)$ random variable  with $\nu=m-3$ degrees of freedom.
Thus, we calculate the probability that $C^2>\chi^2(m-3)$ and if this value is less than 0.05 (a commonly-used significance level), then we reject the null hypothesis that that model is appropriate.
The results of the $\chi^2$ tests are shown in Table II and there is no evidence to reject the null hypothesis in any of the cases.

\section{Response Surface Methods for Agglomeration Distribution}
\label{app:RSP}

Response surface methodology is an experimental design approach usually used for optimizing a process with a response variable that is approximated as a 2\textsuperscript{nd} order function of several input (or factor) variables.
In this work, we use the methodology to understand empirically the effects of film thickness and composition on the negative binomial random variable parameters, $(p,r)$.
We postulate that the following model is appropriate:
\begin{equation}
  \label{eqn:rsp_mod}
  y = \beta_0 +\beta_1 x_1 + \beta_2 x_2 + \beta_{12} x_1x_2 + \beta_{11}x_1^2+\beta_{22}x_2^2
\end{equation}
where $y$ is the response variable ($p$ or $r$), $x_1$ is film thickness and $x_2$ is Ga/(In+Ga).
We employ a face-centered cubic response surface design and the results from each run are shown in Table II (traditionally, two or more replicates of the center point are included---we include only one replicate because of the low variance resulting from simulated, rather than experimental, data).
The parameters ($\beta_i$) were estimated using ordinary least squares and insignificant effects ($p>0.05$, assuming normally distributed error) were removed from the models.
The resulting response surfaces (plotted in Figs. \ref{fig:agglom}e-f) are:
\begin{equation}
  \label{eqn:rsp_fit}
  \begin{split}
    p &= 0.05 + 3.07 x_2 -2.64 x_2^2  \hspace{2.5em} (R^2=0.997)\\
    r &= 0.06 x_1 + 1.12 x_2 - .18 x_1x_2   \hspace{1em}(R^2=0.991)
  \end{split}
\end{equation}

\section{Experimental Methods}
\label{app:exp}
The films analyzed in Section \ref{sec:valid} were produced as follows.
Soda lime glass substrates (2.5~cm~$\times$~2.5~cm) were coated with 700~nm of Mo followed by 700~nm of a Cu-In-Ga mixture using DC magnetron sputtering.
The Cu-In-Ga layers were rotationally sputtered from  Cu\textsubscript{0.8}Ga\textsubscript{0.2} and In sputter targets to yield final Ga/(In+Ga) ratio of 0.25.
The samples were then placed on a graphite sample holder and inserted in a 5~cm diameter quartz reactor tube.
The reactor was evacuated to at least 6~$\times$~$10^{-3}$~Pa to remove gas impurities, and then charged with H\textsubscript{2}Se, H\textsubscript{2}S, and Ar at the desired concentrations.
Samples were heated to 550~\degree C using a 1000~W quartz-halogen lamp and maintained at that temperature for 10~minutes.
Each sample composition was measured using an Oxford Instruments PentaFET 6900 EDX detector in an Amray 1810 scanning electron microscope with the samples in plan view. 
The microscope acceleration potential was set to 20~kV which corresponds to an EDX depth sensitivity of approximately 0.8~\micro m, or one half of the film thickness after reaction.

\section*{References}
\bibliography{refs}
\end{document}